\documentclass[reprint]{revtex4-1}
\usepackage {amssymb, amsmath, graphicx}
\usepackage{gensymb}
\usepackage{amsthm}
\usepackage{amsfonts}
\begin{document}


\title{Possible coexistence of Cycloidal Phases, Magnetic Field Reversal of Polarization and Memory Effect in Multiferroic \emph{R}$_{0.5}$Dy$_{0.5}$MnO$_3$ (\emph{R}=Eu and Gd)} 



\author{Chandan De}
\affiliation{Chemistry and Physics of Materials Unit and International Center for Materials Science,
Jawaharlal Nehru Centre for Advanced Scientific Research, Jakkur P.O., Bangalore 560064 India}

\author{A. Sundaresan}
\email{sundaresan@jncasr.ac.in}
\affiliation{Chemistry and Physics of Materials Unit and International Center for Materials Science,
Jawaharlal Nehru Centre for Advanced Scientific Research, Jakkur P.O., Bangalore 560064 India}


\date{\today}

\begin{abstract}
We report the occurrence of both \emph{ab} and \emph{bc} cycloidal ordering of Mn-spins at different temperatures and their possible coexistence at low temperatures in the polycrystalline mixed rare-earth compounds, \emph{R}$_{0.5}$Dy$_{0.5}$MnO$_3$ (\emph{R} = Eu and Gd), which exhibit extraordinary magnetoelectric properties. While the polarization of Gd$_{0.5}$Dy$_{0.5}$MnO$_3$ is comparable to TbMnO$_3$, the compound Eu$_{0.5}$Dy$_{0.5}$MnO$_3$ shows high value of polarization. However, both of them show giant magnetic tunability  and exhibit large magnetocapacitance whose sign changes across the two cycloidal ordering temperatures. Intriguingly, the electric polarization can be reversed upon ramping up or ramping down the magnetic field, which has not been observed for any of the \emph{R}MnO$_3$ system. Most strikingly, these compounds show non-volatile ferroelectric memory effect even in the paraelectric and paramagnetic region (T$_C$ $\leq$ T $\leq$ 80 K). We attribute these remarkable properties to the coexistence of \emph{ab} and \emph{bc} cycloidal ordered phases.
\end{abstract}


\maketitle 



 Among the spin driven multiferroic materials, the orthorhombic (\emph{o}-) perovskite manganites \emph{o}-$\emph{R}$MnO$_{3}$ ($\emph{R}$ = Gd, Tb, and Dy) have been studied extensively \cite{kimura,spaldin,cheong,yamasaki,kimura anual review,tokura}. In GdMnO$_3$, the cycloidal spins that break the inversion symmetry, are in the \emph{ab} plane and the polarization points along the \emph{a} direction of the orthorhombic (\emph{Pbnm}) structure \cite{noda}. In the manganites with \emph{R} = Tb and Dy, the cycloidal spins are in the \emph{bc} plane and the polarization points along the \emph{c} direction \cite{noda,kimura,goto}. In the case of \emph{o}-$\emph{R}$Mn$O_{3}$ with smaller rare-earths ($\emph{R}$ = Ho, Er, Tm), a collinear  magnetic ordering (E-type) gives polarization along \emph{c}-direction which is substantially higher than that of the $\emph{bc}$ cycloidal phase \cite{ivan,picozzi,lee1,feng}. The existence of different mechanisms of polarization in the $\emph{o}$-$\emph{R}$MnO$_3$ with different \emph{R}-ions indicate that the radius of \emph{R}-ion determines the ferroelectric properties by controlling the competing nearest neighbor ferromagnetic and next-nearest neighbour antiferromagnetic interactions \cite{goto}.

Several studies have been carried out on mixed rare-earth manganites $\emph{R}$$_{1-x}$$\emph{R}$$'$$_{x}$MnO$_{3}$ with $\emph{R}$= Sm, Eu, Tb and $\emph{R}$$'$ = Y and Gd, where multiferroic phases of cycloidal and E-type collinear magnetic structure are found as a function of average radius of rare-earth ions \cite{ishiwata,flynn,tgoto}. Based on magnetic field effects on polarization in Sm$_{0.5}$Y$_{0.5}$MnO$_3$, coexistence of polarization in two different directions has been suggested \cite{fina}. In the case of Dy$_{1-x}$Ho$_x$MnO$_3$, a transition from $\emph{bc}$ cycloidal to $\emph{E}$-type antiferromagnetic phase occurs and coexistence of these two phases is found in a wide compositional range\cite{nzhang}. Application of external pressure in TbMnO$_{3}$ leads to change of $\emph{bc}$ cycloidal ordering to $\emph{E}$-type ordering with a large polarization ($\approx$ 1.0 $\mu$C/cm$^{2}$) \cite{aoyama}.

Here, we report direct evidences for the occurrence of the two cycloidal phases ($\emph{ab}$ and $\emph{bc}$) in two mixed rare-earth manganites, Eu$_{0.5}$Dy$_{0.5}$MnO$_{3}$ (EDMO) and Gd$_{0.5}$Dy$_{0.5}$MnO$_{3}$ (GDMO) and their extraordinary magnetoelectric properties. Contrary to TbMnO$_3$ (TMO), the mixed rare-earth compounds show a large magnetocapacitance accompanied by a change of sign from negative to positive, as a function of temperature. Further, they exhibit large enhancement of electric polarization under applied magnetic field. Surprisingly, they show switching of polarization while ramping up the magnetic field from 0 to 80 kOe. We also demonstrate that the ferroelectric domain state is memorized not only below incommensurate magnetic ordering but also in the paraelectric and paramagnetic region as well.



Details of sample preparation and physical measurements are given in supplemental material {\cite{supp}.  Fig. 1 (a, b and c) shows specific heat divided by temperature (C/T) (left-axis) and magnetization (M) data, measured under field cooled warming (100 Oe) condition (right-axis), with temperature for EDMO, GDMO and TMO, respectively. Since these samples are polycrystalline and the rare-earth moments are higher than the Mn$^{3+}$ moment, we do not observe magnetization anomalies associated with the ordering of Mn$^{3+}$ ions. On the other hand, the (C/T) data clearly show the incommensurate sinusoidal antiferromagnetic ordering (T$_N$), commensurate cycloidal ordering (T$_C$) and $\emph{R}$$^{3+}$ ordering \cite{kimura}, except that the rare-earth moments in EDMO do not order down to 2 K \cite{ordering}. The center panels (d, e and f) show dielectric constant (left-axis) and loss data (right-axis) measured at 50 kHz with different external magnetic fields (0, 40 and 80 kOe). For EDMO and GDMO, the zero field dielectric and loss data show a broad doublet peak which becomes a single peak with a slight positive shift of temperature under applied magnetic field. On the other hand, only a single peak in dielectric and loss data is observed for TMO which becomes broad in presence of magnetic field.

\begin{figure}[t!]
\centering
\includegraphics[width=\columnwidth]{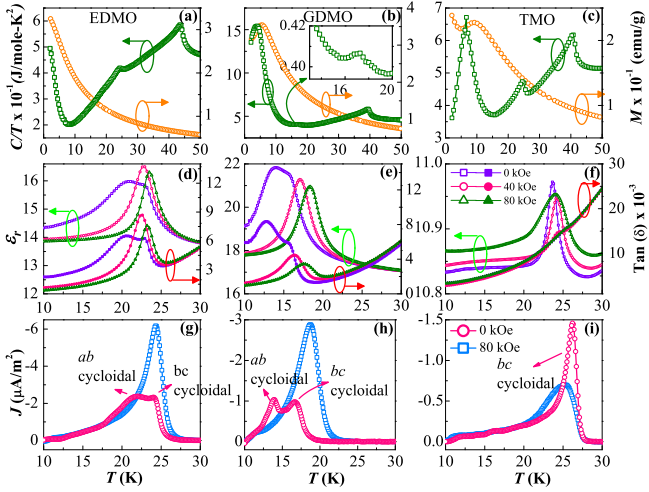}
\caption{\label{fig:sine} First, second and third column represents the data of Eu$_{0.5}$Dy$_{0.5}$MnO$_{3}$, Gd$_{0.5}$Dy$_{0.5}$MnO$_{3}$ and TbMnO$_{3}$, respectively where (Top row) left-axis in a, b and c shows heat capacity divided by temperature and the right axis shows magnetization \emph{vs.} temperature data, (Middle row) the left-axis in d, e and f shows dielectric constant and the right-axis shows loss \emph{vs.} temperature data and (bottom row) g, h and i  shows pyroelectric current \emph{vs.} temperature data.}
\end{figure}

Pyrocurrent data of the samples recorded at 4 K/min, as reported earlier \cite{de}, from 10 to 30 K at 0, and 80 kOe fields, after poling the samples with an electric field (E$_P$ $=$ 8 kV/cm, E$_P$ $\perp$ H$_P$) from 35 to 10 K, are displayed in bottom panels (g, h and i). The warming rate dependent of pyrocurrent confirms the intrinsic ferroelectric nature of the sample \cite{supp}. From these data, we infer that the ferroelectric transition temperatures ($T_C$) for EDMO, GDMO and TMO are 26, 18 and 27 K, respectively. This is in agreement with the heat capacity and dielectric anomaly. It is interesting to note that the zero field pyrocurrent data for both EDMO and GDMO show two peak feature but a single peak for TMO, similar to that observed in dielectric and loss data. From the following discussion, we suggest that the two peak feature in dielectric and pyrocurrent data indicates the presence of $\emph{ab}$ and $\emph{bc}$-cycloidal phases. It is possible that these two cycloidal phases can coexist or there could be a temperature dependent reorientation of cycloidal spins. Considering the average radius of rare earth ions in EDMO and GDMO and phase diagram of temperature versus radius of R-ions, we infer that these two compounds are at the phase boundary between the $\emph{ab}$ and $\emph{bc}$ cycloidal phases \cite{hemberger}. In agreement with the earlier report \cite{nabe1}, thermal hysteresis observed around the dielectric anomalies confirms that the phase transition between these two cycloidal phases is first order which indicates possible coexistence of these phases \cite{supp}. However, it requires a single crystal study to confirm the phase coexistence. Based on the theoretically obtained magnetoelectric phase diagram of temperature versus $\emph{J}$$_2$ (Next nearest neighbour exchange interaction), we attribute the low temperature (LT) peak to $\emph{ab}$ and high temperature (HT) peak to $\emph{bc}$ cycloidal ordering \cite{mochizuki}. Under an applied magnetic field (80 kOe), the two peak feature disappears and becomes a single peak with enhanced pyrocurrent and significant increase of T$_C$. The disappearance of LT peak indicates conversion of $\emph{ab}$ cycloidal into $\emph{bc}$ cyclodial phase which is consistent with the enhanced pyroelectric current. In contrast, the magnitude of the single peak observed for TMO is decreased with magnetic field without any significant change in T$_C$ [Fig.1 (i)]. The broad nature of the current at 80 kOe may indicate the partial conversion of \emph{bc} cycloidal phase into \emph{ab} cycloidal. It should be noted that the polarization in GdMnO$_3$ is suppressed strongly with applied magnetic field through weakening of the commensurate cycloidal ordering \cite{zhang,supp}.
\begin{figure}[b!]
\centering
\includegraphics[width=\columnwidth]{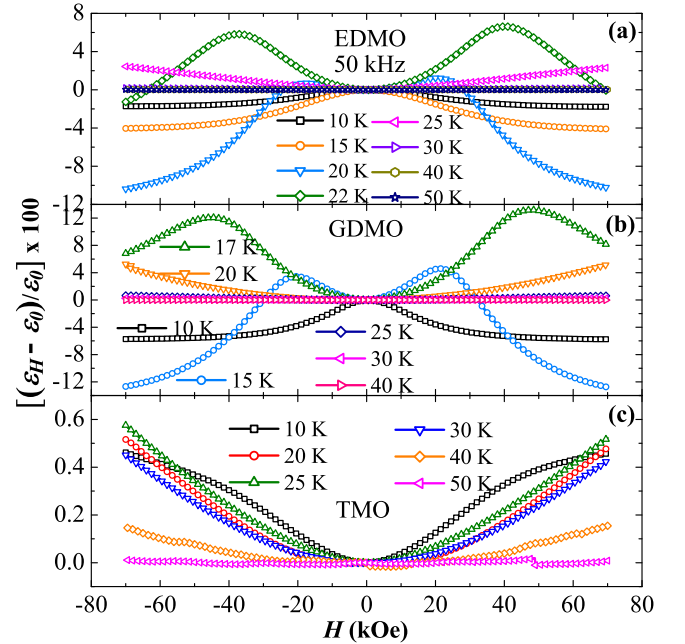}
\caption{\label{fig:sine} Magnetocapacitance data of (a) Eu$_{0.5}$Dy$_{0.5}$MnO$_{3}$, (b) Gd$_{0.5}$Dy$_{0.5}$MnO$_{3}$ and (c) TbMnO$_{3}$ measured at 50 kHz at various temperatures.}
\end{figure}

Fig. 2 (a, b and c) shows magnetocapacitance (MC) data measured at 50 kHz while sweeping the magnetic field from -70 kOe to +70 kOe at the rate of 100 Oe/sec at different temperatures for all the three samples. The data measured at 500 Hz and 2 MHz are shown in supplemental material \cite{supp}. It is intriguing to note that the behavior of MC in EDMO and GDMO is quite different from TMO. Above T$_C$, all the three samples exhibit positive MC and it reaches a maximum around T$_C$. Below T$_C$, the MC in TMO remains positive and at 10 K it levels off above 40 kOe as shown in Fig. 2c. In contrast, the mixed rare earth samples show a crossover from a positive to negative MC on decreasing the temperature from paraelectric to ferroelectric state through an intermediate temperature range (between the two pyrocurrent peaks) where the positive MC shows a broad maximum, corresponding to a critical field, which shifts to lower field with decreasing temperature. The existence of such critical field indicates the change in direction of polarization \cite{goto}. Below certain temperature, MC becomes completely negative. This behaviour is consistent with the fact that MC is positive below the cycloidal ordering temperature in TbMnO$_3$ and negative in GdMnO$_3$ \cite{supp}. It is also important to note the large MC (12\%) observed in wide temperature range for the mixed rare-earth compounds compared to that in TMO (0.6\%).

Fig. 3(a, b and c) shows temperature dependent polarization data, at different magnetic fields, obtained by integrating the pyrocurrent (inset), recorded after poling the samples with E$_P$ $=$ 8 kV/cm $\perp$ H$_P$. In TMO, it is known that the polarization changes its direction from $\emph{c}$- to $\emph{a}$-axis when magnetic field is applied along the $\emph{b}$-direction \cite{kimura,kimuragoto}. However, the observed polarization along $\emph{a}$-direction is smaller because of the lack of complete flipping of $\emph{bc}$ cycloidal phase due to high domain wall formation energy which is determined by the competition between Zeeman energy and magnetic anisotropy\cite{murakawa,kagawa,abe}. In the present case, the polycrstalline TMO shows only a small decrease in polarization ($\Delta P$ $\approx$ $-$12\% at 80 kOe) with magnetic field as shown in inset of Fig. 3c. In contrast, a dramatic change of polarization is observed in the two mixed rare-earth manganites, as shown in the insets of FIg. 3a and 3b. The magnitude of the polarization at zero field is fairly large (almost four times that of TMO) in EDMO and it increases with magnetic field drastically as shown in the inset of the fig. 3a. Remarkably, the effect of magnetic field on polarization in GDMO is very large ($>$140\% at 80 kOe) (inset of fig. 3b ) although its zero field value is comparable to TMO. The evolution of pyrocurrent peak upon applied magnetic field is shown in right hand side insets. It is clear from these figures that the two peak nature gradually becomes a single peak both in EDMO and GDMO indicating the conversion of LT cycloidal phase into HT cycloidal phase. It is also interesting to note that the magnitude of pyrocurrent and T$_C$ increases with magnetic field. We suggest that the large enhancement of polarization is due to the coexistence of $\emph{ab}$ and $\emph{bc}$-cycloidal phase at 0 kOe and change of $\emph{ab}$ cycloidal into $\emph{bc}$ cycloidal phase in applied magnetic field. At zero field, we observe a net polarization of the two components i.e. along $\emph{a}$ and $\emph{c}$ directions. Besides, we propose that the $\emph{bc}$-cycloidal regions can act as seed for changing the rotation plane of $\emph{ab}$ cycloidal so that the rotation of cycloidal plane becomes easier.
\begin{figure}[t!]
\centering
\includegraphics[width=\columnwidth]{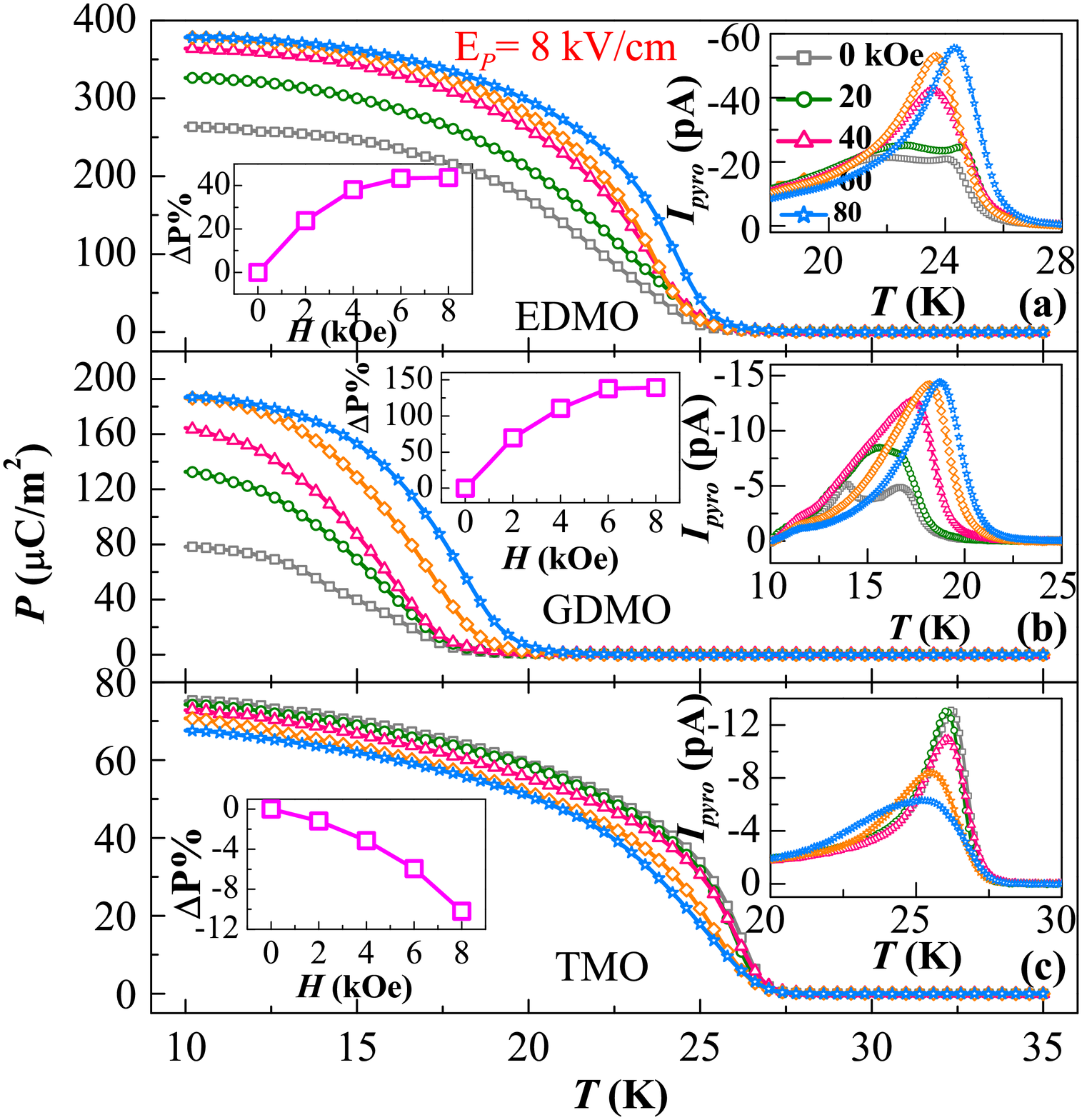}
\caption{\label{fig:sine} Polarization data of (a) Eu$_{0.5}$Dy$_{0.5}$MnO$_3$, (b) Gd$_{0.5}$Dy$_{0.5}$MnO$_3$ and (c) TbMnO$_3$ measured at various magnetic field. Inset (right side) shows pyrocurrent data. Inset (left side) shows normalized $\bigtriangleup$P $=$ [(P(H)$-$P(0)/P(0))$*$100] change of polarization $\emph{vs.}$ magnetic field data.}
\end{figure}
\begin{figure}[b!]
\centering
\includegraphics[width=\columnwidth]{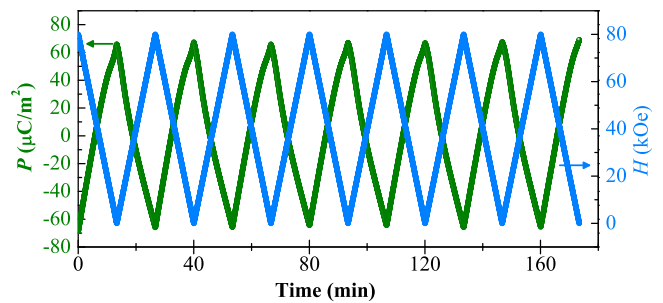}
\caption{\label{fig:sine} Periodic change of polarization (left-axis ) and magnetic field (right-axis) with time recorded at 7 K after poling the sample from 30 to 7 K with $-$8 kV/cm and 80 kOe in GDMO.}
\end{figure}

Fig. 4 shows the polarization obtained by integrating the zero bias magnetoelectric current measured isothermally at 7 K by sweeping the magnetic field at 100 Oe/sec from 80 to 0 to 80 kOe for six cycle after magnetoelectric poling with E$_P$ $=$ $-$8 kV/cm $\perp$ H$_P$ $=$ 80 kOe from 30 to 7 K for GDMO. We see a sequential flipping of polarization without any decay from positive to negative and negative to positive upon ramping up the magnetic field from 0 to 80 kOe and then ramping down to 0 kOe. Though similar switching behavior is reported in other multiferroics \cite{lee,hur,Kouji,Yamasaki}, it is not known in {\it R}MnO$_3$. The observed polarization reversal in the present case is explained by the sequential conversion of the cycloidal phases. As we poled the sample with H$_P$ $=$ 80 kOe, the polarization at 7 K should be directed along $\emph{c}$-direction (only $\emph{bc}$ cycloidal phase). Upon ramping down the field to zero, $\emph{ab}$ cycloidal phase grows and the polarization flips 90$\degree$ ($\emph{a}$-direction) and the net polarization decreases to zero at H $\sim$40 kOe where the oppositely aligned domains (along $\emph{c}$-direction) are equally populated. When the field is further ramped down to zero, the polarization goes to the opposite (negative) direction. A complete switching of polarization with equal magnitude is obtained for a field range of 0 to 80 kOe. The actual mechanism of switching of polarization depends on the orientation of neighbouring \emph{ab} and \emph{bc} cycloidal phases.
\begin{figure}[t!]
\centering
\includegraphics[width=\columnwidth]{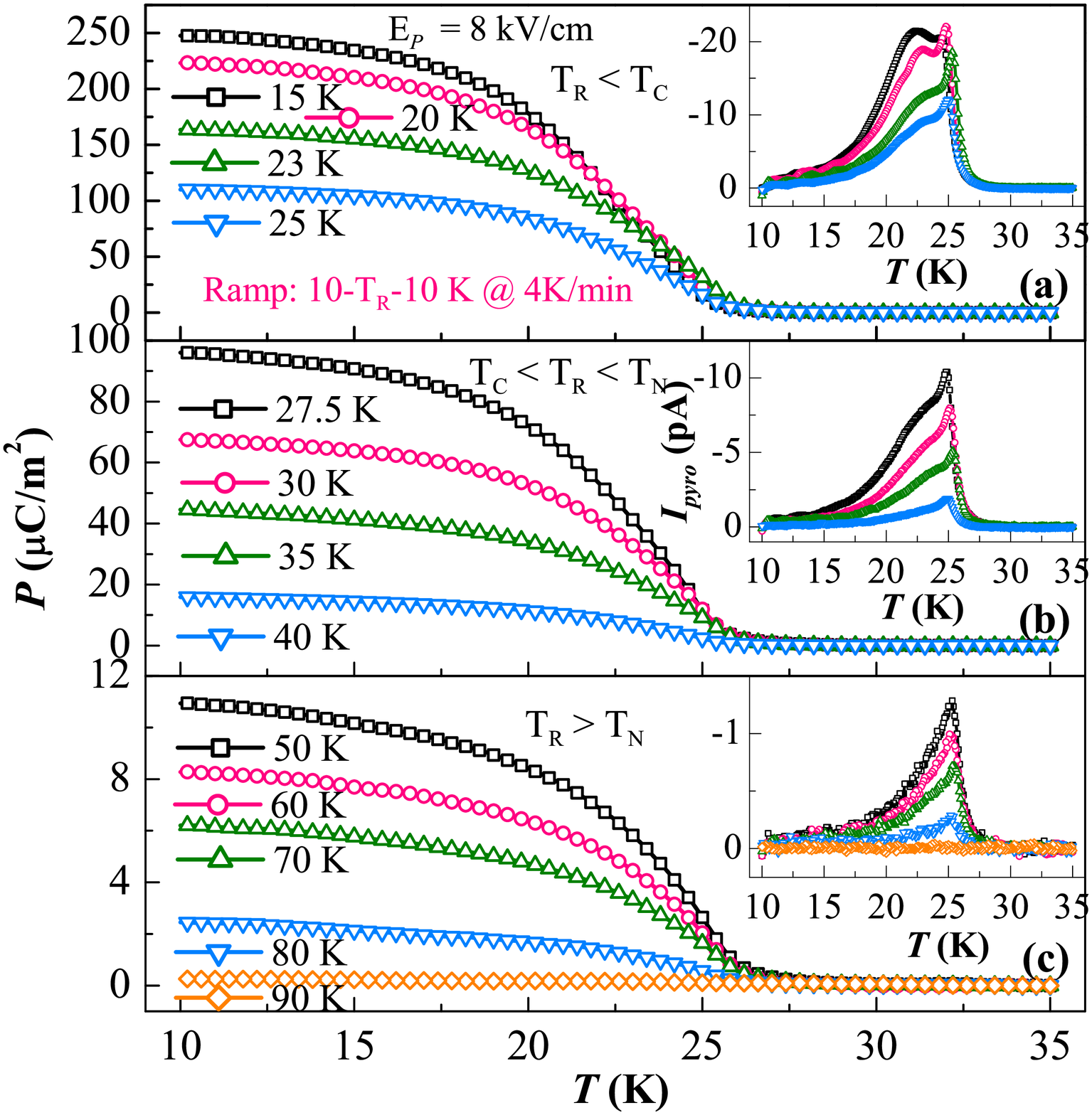}
\caption{\label{fig:sine} Polarization data after integrating the pyrocurrent (inset), recorded after poling the sample (EDMO) from 30 to 10 K with 8 kV/cm, 0 kOe and ramp the temperature from 10 to T$_R$ to 10 K with zero bias.}
\end{figure}

 Finally, we present the observation of memory effect, the retention of polarized state when the sample is warmed to the paraelectric state \cite{fina,taniguchi,finger}.  In this memory experiment, first the sample was poled (E = 8 kV/cm) from 30 K to 10 K. After poling the sample, the electrode wires were short-circuited and the sample was warmed up to a temperatures, called ramping temperature (T$_R$) and again cooled down to 10 K. After this ramping treatment, the pyrocurrent was recorded from 10 to 35 K at 4 K/min rate (see the inset of Fig. 5). Fig. 5a shows the integrated polarization versus temperature for T$_R$ $<$ T$_C$ (T$_R$ $=$ 15, 20, 23 K). As expected the polarization is decreased gradually with increasing T$_R$. Fig 5 (b) shows the polarization for T$_C$ $<$ T$_R$ $<$ T$_N$ (T$_R$ $=$ 27.5, 30, 35 and 40 K). Though the polarization decreases with increasing T$_R$, it is interesting to note that the polarization survives even after warming the sample above T$_C$. Fig 5(c) shows the polarization data obtained from similar measurement protocol for T$_R$ $>$ T$_N$ (T$_R$ $=$ 50, 60, 70, 80 and 90 K). Surprisingly, we see that the polarization is memorized up to 80 K and vanishes at 90 K. To observe the memory effect above T$_N$, it is very important to pole the sample just above the cycloidal ordering temperature to avoid the formation of internal electric field \cite{de}. In fact, we have shown that if we pole the sample from slightly high temperature (50 K) we do not see the memory effect above T$_N$ \cite{supp}. It is because of the poling temperature, the memory effect is not seen in Sm$_{0.5}$Y$_{0.5}$MnO$_3$ \cite{fina}. The variation of polarization obtained at 10 K as a function of T$_R$ reveals the existence of three different slopes that correspond to temperatures below T$_C$, T$_C$ $\leq$ T $\leq$ T$_N$ and above T$_N$ \cite{supp}. To further confirm the memory effect, we poled ($-$8 kV/cm) the sample only once from 30 to 10 K and measured the pyroelectric current continuously while warming and cooling from 10 K to different T$_R$. In this measurement, we observed depolarization and polarization current in each warming and cooling cycle for T$_R$ up to 80 K\cite{supp}. Similar memory effect is also observed in GDMO \cite{supp}. These results demonstrate the presence of memory effect(cycloidal phase) at temperatures much above the ferroelectric transition. From the inset of Fig. 5, we see that the two shoulders in the pyrocurrent peak, which are indicative of  $\emph{ab}$ and $\emph{bc}$ cycloidal phases, gradually becomes one where the LT peak disappears with increasing T$_R$. This result suggests that the $\emph{bc}$ cycloidal phase (HT peak) is responsible for the memory effect. However, we suggest that this finding requires further study of dielectric response to explain how the cycloidal phase exists in the paramagnetic region also \cite{taniguchi}.

In conclusion, we have shown the possible coexistence of $\emph{ab}$ and $\emph{bc}$ cycloidal phases in the mixed rare-earth multiferroic manganits, Eu$_{0.5}$Dy$_{0.5}$MnO$_3$ and Gd$_{0.5}$Dy$_{0.5}$MnO$_3$. As a result, these materials exhibit large magnetic tunability of polarization and high magnetocapacitance. More importantly, the electric polarization can be switched by ramping the magnetic field. Further, the electric polarization retains its memory even in the paraelectric and paramagnetic region. We suggest that these effects results from the coexistence of cycloidal phases.

The authors acknowledge the Sheikh Saqr Laboratory at the Jawaharlal Nehru Centre for Advanced Scientific Research for experimental facilities.



%
%

%



\begin{thebibliography}{0}
\bibitem{ordering} In contrast to DyMnO$_3$, the magnetic ordering temperature in EDMO is decreased because of the fact that the magnetic Dy$^{3+}$ ions are diluted by the non magnetic Eu$^{3+}$ ions.
\end{thebibliography}

\end{document}